\documentclass[11pt]{article} 
\input epsf.tex

\newcommand{\be}{\begin{equation}}
\newcommand{\ee}{\end{equation}}
\newcommand{\bea}{\begin{eqnarray}}
\newcommand{\eea}{\end{eqnarray}}

\def\grad{\nabla}

\def\p{\partial}

\newcommand{\hc}{\ensuremath{h_{\mathrm{crit.}}}}

\begin{document}



\vspace{1cm}

\begin{center}
{\large\bf Density matrix renormalization group analysis of the spin $\mathbf{1/2}$ XXZ chain 
in an XY symmetric random magnetic field}
\end{center}
\vspace{5mm}

\begin{center}

{\large Laura Urba$^{\scriptstyle 1}$ and Anders Rosengren$^{\scriptstyle 2}$. \\ }

\vspace{5mm} 
KTH-SCFAB, Department of Physics, Condensed matter theory, SE-106 91 Stockholm, Sweden.

\vspace{3mm}

\vspace{5mm}

{\tt
$^1$ laura@condmat.physics.kth.se \\
$^2$ ar@condmat.physics.kth.se
}

\end{center}

\vspace{5mm}

\begin{center}
{\large \bf Abstract}

\end{center}
\noindent

 The spin $\frac{1}{2}$ $XXZ$ chain in a random magnetic field pointing in the $Z$ direction is numerically studied
using the Density Matrix Renormalization Group (DMRG) method. The phase diagram as a function of
the anisotropy of the $XXZ$ Hamiltonian and the strength of the random field is analyzed by computing the spin 
correlations and the superfluid density. To obtain the superfluid density we consider a superblock 
configuration representing a closed system with an arbitrary twist at the boundary. This allows 
us to estimate the size of the critical region where the quasi-long-range order persists, that 
is, where the spin correlation length is infinite and the superfluid density is non-zero.

\vspace{2cm}
\begin{flushleft}
PACS \hspace{0.2cm} 75.10.Jm, 75.30.Kz, 75.45.+j.
\end{flushleft}
\vfill
\begin{flushleft}
October 2002
\end{flushleft}

\section{Introduction}
The study of quantum phase transitions represents a wide area of research 
in condensed matter physics~\cite{review}.
These transitions occur at $T=0$ when a parameter of the hamiltonian is 
varied across
a critical value and are driven by quantum fluctuations. Among them, those 
that are due to the effect of disorder 
are more difficult to analyze both analytically and numerically and have been 
the focus of much attention recently.  

In this paper we consider the spin $1/2$ XXZ chain in the presence of a 
non-symmetry breaking random
magnetic field. 
This system is particularly interesting since a renormalization group 
analysis \cite{giamarchi,nagaosa,fisher} 
suggests the existence of a critical region in the space of parameters 
where the disorder is not relevant. This idea is also supported by numerical 
calculations \cite{runge,WSJP}. In \cite{runge} the system was studied using 
an exact diagonalization method up to $16$ sites. Although this size is enough 
to study the system far from the critical region, near the transition, as 
it is explained there, the calculations were less reliable due to the small size of the 
chains considered.
 Larger chains can be handled using a density matrix renormalization group 
(DMRG) algorithm \cite{white} at the expense of truncating the Hilbert space 
of states. 
 This method was used in \cite{WSJP} to obtain further information about
the critical region by computing the phase sensitivity, which is the difference 
in energy between periodic and antiperiodic boundary conditions. 

In this paper we use the DMRG method to compute the correlation length
and the superfluid density which give direct information about the existence
of a quasi-long-range order region since that region is defined as the one
where the correlation length diverges and the superfluid density is non-zero. 
 In particular, to compute the superfluid density we used a superblock 
configuration that allowed us to introduce an arbitrary twist at the boundary.

Since we perform DMRG calculations, our results contain two approximations. 
The first is that we consider a finite chain
and the second that we truncate the Hilbert space (keeping a finite number of 
density matrix eigenstates). These approximations are related since with a 
truncated Hilbert space the correlation length is always finite and, in fact, 
the value corresponding to an infinite chain (what we are going to call the 
'truncated system correlation length') becomes smaller as we reduce the 
Hilbert space.     
To minimize the finite size effects in the calculation of the spin correlations,
 it is necessary to consider chains larger than the truncated system 
correlation length. However, in the critical region this correlation length 
increases rapidly as we keep more DMRG states, therefore, reducing the 
truncation error by keeping a large number of states soon becomes prohibitive 
since we need to consider very long chains \footnote{For a more quantitative 
statement see Fig. \ref{clevsL}}. Thus, most of 
our calculations of the spin correlations are done keeping 
up to 32 states and give only an estimation to the shape of the critical region. 
As we will discuss later, the main source of error in this 
estimation is the truncation of the Hilbert space. To further support these 
qualitative results, for the special case of anisotropy $\Delta=-0.75$ and selected 
values of the random magnetic field, we keep up to 96 states and extrapolate the 
computed correlation length to infinite number of sites and infinite number of 
DMRG states. The extrapolations allow us to conclude the existence of a 
quasi-long-range order region and put bounds on the maximum critical field. 

In the calculations of the superfluid density we also keep up to 32 states in 
most of the cases and estimate the shape of the critical region using a 
criterion in which, again, the main approximation is the truncation of the 
Hilbert space. As we did for the 
correlations, for the special case of $\Delta=-0.75$ and selected values of the random 
magnetic field we keep up to 128 states and extrapolate the superfluid density to 
infinite length and infinite number of DMRG states. We find an agreement between the 
correlation length and the superfluid density results.  

The remainder of this paper is organized as follows. In section~\ref{bkgsec} 
we summarize the known results for this system and in section~\ref{nummeth} we 
describe the numerical method used. We present and discuss
the results in sections~\ref{ressec}~and ~\ref{extrapol} and finally, in 
section~\ref{concsec} we give our conclusions.

\section{The spin $1/2$ $XXZ$ chain in a random $z$ magnetic field} \label{bkgsec}

 In this section we summarize previous results obtained for this system by C. Doty and D. Fisher~\cite{fisher}.

 In the absence of disorder the spin $1/2$ XXZ chain is defined by the 
Hamiltonian  
\be
H_0 = \sum_{i=1}^L (S_i^x S_{i+1}^x +S_i^y S_{i+1}^y + \Delta S_i^z S_{i+1}^z).
\label{eq:H0}
\ee
When the anisotropy parameter, $\Delta$, satisfies $-1 < \Delta \le 1$, the system is in a non-magnetic gapless phase 
where the spin-spin correlation functions in the ground state decay as a power law (quasi-long-range order (quasi-LRO) 
phase) \cite{affleck,mattis,young}. 
On the other hand, the spectrum has a gap for $\Delta > 1$ ($\Delta \le -1$) and possesses a long-range Ising 
like antiferromagnetic (ferromagnetic) order.

In this work we study the effects produced by the addition of the following term:

\be
H_1= \sum_{i=1}^L h_i^z S_i^z,
\label{randf}
\ee

where $h_i^z$ is a random magnetic field with an independent and uniform
distribution in each site
($\langle h_i^z \rangle=0$, $\langle h_i^z h_j^z \rangle =\delta_{ij} h_0^2 /3$).
Note that the addition of this term does not break the symmetry in the $xy$ plane.

The main conclusion of \cite{fisher} is that for $-1< \Delta < -1/2 $ the  quasi-long-range order that 
the system possesses at $h_z^i=0$ persists for finite values of the random magnetic field. 

 To reach this conclusion, the long-distance behavior of the system was analyzed by studying the
renormalization group flow. In the region $|\Delta | < 1$ one can perform a Jordan-Wigner transformation to
spinless fermionic variables. The component $S^z$ is converted into the fermion occupation number and $S^\pm$
are creation and annihilation operators:
\bea
S^z_n &=& \Psi_n^\dagger \Psi_n - \frac{1}{2}\ , \\
S^-_n &=& \exp\left(i\pi\sum_{m<n}(S^z_m + \frac{1}{2})\right) \Psi_n .
\eea
For the pure XXZ chain when $\Delta=0$, the ground state is a half-filled band with Fermi momenta $k_F=\pm\pi/2a$ where $a$ is 
the lattice spacing. Low energy excitations around the Fermi surface can be described in terms of bosonic fields using a standard 
procedure~\cite{haldane}:
\be
\Psi_n = \sqrt{\frac{a}{L}}\left[ e^{-ik_Fx-i \Phi_R(x)} + e^{ik_Fx+i\Phi_L(x)}\right].
\ee
Here $L$ is the length of the system. The bosonic fields $\Phi_{L,R}$ describe left and right moving excitations. 

The random perturbation (\ref{randf}), can also be expressed in terms of the bosonic fields and, at low energies, 
the continuum limit can be taken which results in an action~\cite{giamarchi,fisher}:
\be
S = \int dx d\tau ~\left( \frac{1}{2} \kappa (\p_{\tau}\Phi)^2 +\frac{1}{2} \kappa (\p_{x}\Phi)^2 +
g ~ \cos(2\Phi) + \eta (x) \p_{x}\Phi + \rho (x) e^{i\Phi} + \rho^*(x) e^{-i\Phi} \right)\ ,
\ee
for the field $\Phi(x) = \Phi_R + \Phi_L$. Besides, $g$ is a function of the anisotropy
$\Delta$, and $\eta(x)$ and $\rho(x)$ are composed of the Fourier components of the random field near zero and $2k_F$, respectively.
 Also, $\kappa$ is the spin stiffness of the ordered system given by the Bethe-ansatz solution \cite{baxter}:
\be
\kappa = \frac{1}{2 \pi} \left( 1-\frac{1}{\pi} \cos^{-1}
\Delta  \right).
\ee
Afterwards,  the replica formalism can be used to compute disorder averages of ground-state expectation values
giving the effective action:
\bea
\overline{S_{eff}} &=& \sum_{\alpha =1}^n \int dx~d\tau~ \left[ \frac{1}{2} \kappa (\p_{\tau} \Phi_{\alpha})^2 +
\frac{1}{2} \kappa (\p_x \Phi_{\alpha})^2 \right] -  \nonumber \\
 && - \sum_{\alpha ,\beta} D \int dx~d\tau~d\tau'~
\cos[\Phi_{\alpha}(x,\tau) - \Phi_{\beta}(x,\tau')],
\label{eq:action}
\eea
where $D=<(h_i^z)^2>=h_0^2 /3$. 
Thus, the disorder induces interactions between pairs of replicas. The interaction comes from 
averaging over $\rho$, that is the term due to the staggered component of the magnetic field. The 
term due to $\eta$ is irrelevant at large distances and is not included in (\ref{eq:action}).

Using the fact that the dimension of the operator $\cos(\Phi)$ is $1/4\pi\kappa$,  the RG eigenvalue $\lambda$, 
which determines the rescaling of $D$ is given by \cite{fisher}:
\be
\lambda = 1+2z-2\zeta,
\ee
where the dynamical exponent $z=1$ and $\zeta=1/(4\pi\kappa)$.

The sign of $\lambda$ determines the relevance of the randomness $D$. In the interval $-1/2 < \Delta \le 1$,
$\lambda$ is positive, indicating that the disorder is relevant and that the quasi-LRO is unstable under any small 
amount of randomness. On the other hand, $\lambda$ is negative in the region $-1 < \Delta < -1/2$, so there 
the disorder is irrelevant and the quasi-LRO is expected to persist for small values of $D$. Therefore, in the 
latter case, a quantum phase transition between a quasi-LRO phase and a localized disordered phase has to 
occur for finite $D$~\cite{fisher}. By means of a RG treatment for weak disorder \cite{giamarchi} the RG equations governing the 
critical properties can be obtained \cite{fisher}. The flow lines in the $\kappa - D$ plane, near the point 
$\kappa=1/6 \pi$ are parabolas, $(\kappa-1/6 \pi)^2 - D = 0$ being the critical one. The phase transition is of 
the Kosterlitz-Thouless type \cite{fisher} and the correlation length corresponding to the spin correlation functions 
diverges like:
\be
\xi \sim \exp (A/ \sqrt{\Delta-\Delta_c}),
\ee
when the quasi-LRO phase is approached along a path in the phase diagram at fixed disorder.
In the region where the pure system is an Ising like ferromagnet (antiferromagnet) the Imry-Ma argument  
~\cite{ma} can be used to conclude that an arbitrarily weak random $z$ field will destroy the long-range order. 

\begin{figure}
\centering
\epsfysize=8.0truecm
\epsffile{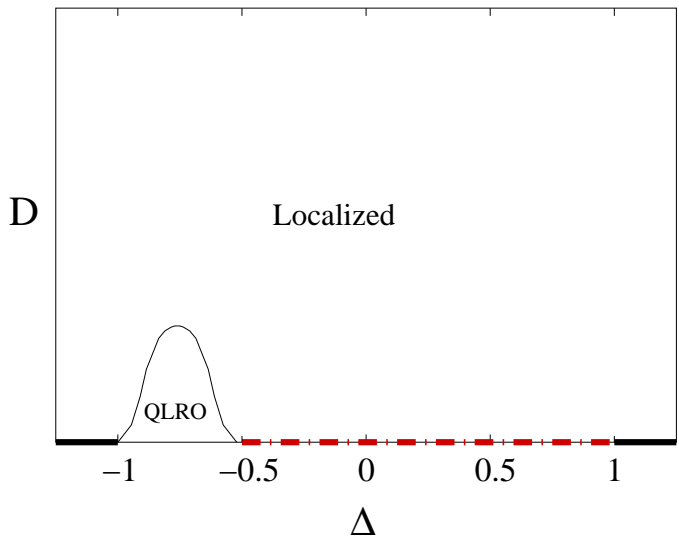}
\caption{Phase diagram corresponding to the XXZ chain in a random field in the $z$ direction according to \cite{fisher}.}
\label{phasediag1}
\end{figure}

The phase diagram $h_0-\Delta$ proposed by Doty and Fisher is shown in Fig. \ref{phasediag1}. Note that the 
top of the region where the quasi-LRO persists i.e the maximum critical field is not established since the perturbative renormalization group result is valid only 
near $\Delta=-0.5$. 

A numerical work analyzing this phase diagram was performed by Runge and Zimanyi \cite{runge}. They use an exact 
diagonalization procedure up to $16$ sites and compute, among other things, the superfluid density $\rho_s$ which 
measures the sensitivity of the system to changes in the boundary conditions: 
\be
\rho _s = \frac{1}{L} \frac{\partial ^2 E}{\partial (\grad \theta)^2}.
\label{rhoeq}
\ee
Here, $\theta$ is a twist introduced at the boundary by adding to the Hamiltonian (\ref{eq:H0}) the term
\be
\frac{1}{2} \left( S_1^+ S_L^- e^{i\theta} + S_1^- S_L^+ e^{-i\theta} \right) + \Delta \, S_1^z S_L^z.
\label{eq:twisted}
\ee 
Doing this, the chain is closed in such a way that a boson acquires a phase $e^{i\theta}$ ($e^{-i\theta}$) if 
it is hopping to the right (left) through the boundary. 
  
This superfluid density is the order parameter of the transition ($\rho_s > 0$ in the quasi-LRO phase 
and $\rho_s=0$ in the localized phase). 
 
In \cite{runge}, the authors find a qualitative agreement with the RG predictions 
\cite{giamarchi,fisher,nagaosa}. However they conclude that it is necessary to study larger chains to obtain more accurate results near the critical region. 

In this paper we compute the superfluid density and the correlation length using the DMRG approach which allows us to 
study longer chains. In that way we expect to obtain further evidence and complementary results.
    
\section{Numerical method} \label{nummeth}

We compute the properties of the system using an infinite-size DMRG algorithm \cite{white}. The disorder is included by doing the calculations for a single configuration of the random field and then, averaging over many realizations.  
Since the ground-state is non-magnetic in the quasi-LRO regime for the pure system and is expected to be also non-magnetic in the whole range of $\Delta$ when the disorder is turned on, we perform our calculations constrained to the Hilbert 
subspace where the $z$-component of the total spin is zero. We have checked that this approximation is justified after sufficient averaging is performed.   

We calculate the spin correlations 

\be
C_j(r)=\langle S_0^x S_r^x + S_0^y S_r^y \rangle,
\label{correlation}
\ee

using open boundary conditions and the same superblock configuration $B \bullet \bullet B$ as in \cite{anxy} 
(where $B$ represents a block and $\bullet$ a single site). In (\ref{correlation}), $\langle \ldots \rangle$ 
denotes quantum expectation value in the ground state and $j$ is a given realization of the random field. 
 To avoid the boundary effects, we first grow the system until the length of each block is around 30 sites.
After that we start to compute the correlations taking the spin operator $S_0^+$ corresponding to the left 
site added at that DMRG step. In the next 
iteration this operator is rotated and truncated using the eigenvectors of the density matrix and so 
it will correspond to the right-most site of the left block. Iterating this procedure for the same 
operator we obtain after $r$ DMRG steps an operator $S^+_r$ associated to a site which is at a distance $r$ from the 
center. Simultaneously, the correlation $C_j(r)$ is computed at each of those DMRG steps  as 
$\langle S^-_0S^+_r + S^+_0 S^-_r \rangle$. Using this standard procedure it is only necessary to store 
one spin matrix ($S^+_r$).
Afterwards, we average the logarithm of the correlation functions $C_j(r)$ over disorder to obtain what is 
called  the typical correlation \cite{fisher2}. Usually, more than 150 configurations of the random field are required. 

At large distances $C_j(r)$ should be exponential outside the critical region 
where the disorder is relevant (finite correlation length) and power law 
inside the critical region, where it is irrelevant (infinite correlation length). 
However, due to the truncation of the DMRG and the finite size effects, we also find an exponential behavior inside the critical region. 
This fact makes possible to define a typical correlation length from the slopes of the curves $\ln C_j (r)$ vs $r$ and analyze its behavior in the phase diagram 
$\Delta-h_0$. The correlation length computed in this way is of course  
approximated and consequently gives qualitative information about the phase diagram. 

If chains of various length are considered and, if for each of them, various number 
of states are kept, an extrapolation to infinite length and infinite number of DMRG 
states can be performed to decrease the error. This is discussed in section 
\ref{extrapol} where we do the extrapolations for a few special points in 
the phase diagram.    

\begin{figure}
\centering
\epsfysize=6.0truecm
\epsfbox{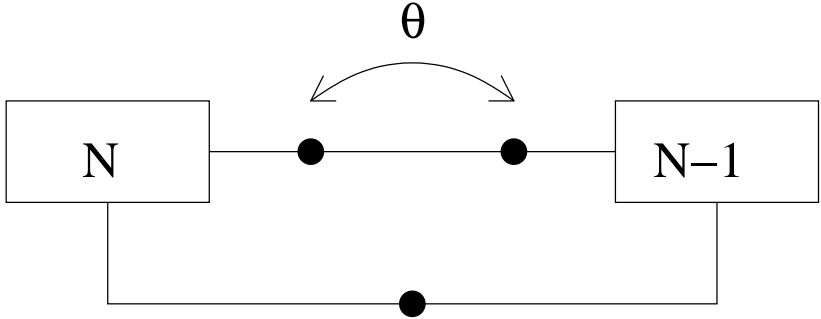}
\caption{Superblock configuration used for the calculation of the superfluid density. 
The twist is introduced between the two single sites added at each step of the DMRG.}
\label{superblock}
\end{figure}

To compute the superfluid density which is the order parameter of the transition 
between the quasi-LRO and the localized states, we need to consider a closed 
system with twisted boundary conditions.
Closed systems can also be treated with the density matrix algorithm as shown in~\cite{white2}. In that work two superblock configurations were analyzed: $B \bullet B \bullet$ and $B \bullet \bullet B$. 
It was found that the first one is better since it has a smaller truncation
error (because the two blocks are not connected directly), although larger than the error obtained with open boundary 
conditions. In our case, we found the configuration $B \bullet \bullet B$  more convenient to introduce twisted boundary conditions 
since the twist can be introduced between the two single sites that we add at each step of the procedure (in that case 
the operators $S^+$ and $S^-$ which appear in (\ref{eq:twisted}) are not truncated). 
On the other hand since we are also interested in minimizing the truncation error, for the actual calculation 
we chose the configuration in Fig. \ref{superblock} which combines the advantages of both.  
Note that the left and right blocks have different number of sites although the same number of states are 
kept on both of them. This asymmetry implies that we have to compute two density matrices, one for the left 
block and one for the right one. However this is necessary anyway since the reflection symmetry is already
broken by the random magnetic field. 

To approximate eq. (\ref{rhoeq}) numerically, given a configuration of the random magnetic field, we perform the DMRG 
procedure twice (with and without twist) and compute for each length $L$ the finite difference
\be
\rho _s (L) \simeq \frac{1}{L} \frac{E(L \grad \theta)-E(0)}{(\grad \theta)^2} ,
\label{rhocal}
\ee   
 where we used that the first derivative $\partial E/\partial \nabla \theta$ vanishes at $\nabla \theta =0$ due to the
symmetry  $E(\nabla \theta) =  E(-\nabla \theta)$. 
 Afterwards, we average this quantity over many configurations of the field. The h
larger the strength of the field the more configurations needed (in most of 
the cases more than 200 configurations were considered).   

\begin{figure}
\centering
\epsfysize=8.0truecm
\epsfbox{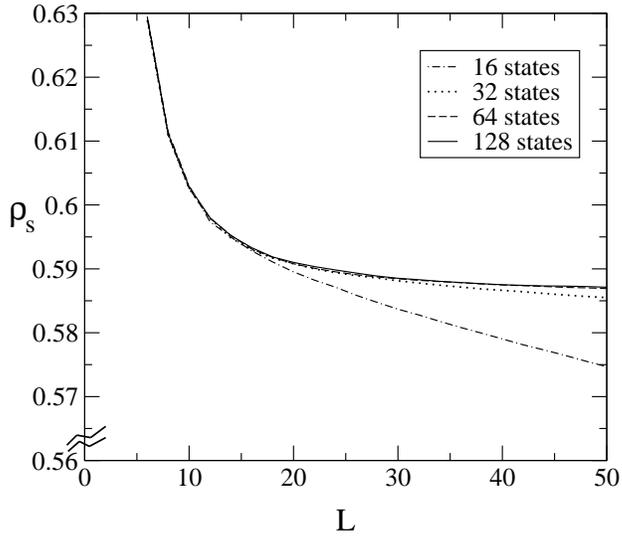}
\caption{The dependence of the superfluid density with the length of the chain for $\Delta=-0.75$ 
and a random field with strength $h_0=0.1$ keeping 16, 32, 64 and 128 eigenstates of the density matrix.  
It is possible to observe the convergence of the method.}
\label{difstates}
\end{figure}

To test the convergence of the DMRG method we computed $\rho_s$ keeping different number of eigenstates of the density matrix. It was found that the DMRG procedure converges if we keep $\grad \theta$ constant i.e. the twist 
$\theta$ increases with the length of the chain. This is shown in Fig. \ref{difstates}, where $\rho_s$ vs. $L$ 
for $\Delta=-0.75$ is plotted for a random field $h_0=0.1$ keeping up to 128 eigenstates of the density matrix.  
Since we need $L \nabla \theta$ to be small, in 
the calculations we  took $\grad \theta=0.01$ and grew the system up to $50$ sites.

\begin{figure}
\centering
\epsfysize=8.0truecm
\epsffile{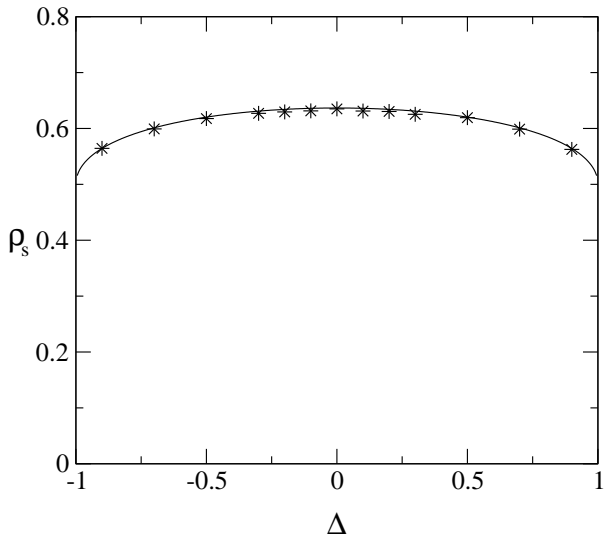}
\caption{The dependence of the superfluid density with the anisotropy for the system without disorder. Our numerical results for a chain of 50 sites keeping 32 eigenstates of the density matrix are compared to the exact solution found in \cite{young}.}
\label{comparho}
\end{figure}

We also checked (see Fig. \ref{comparho}) that without disorder our numerical results for $\rho_s$ 
agree with the exact solution 
\be
\rho_s (\Delta , h_0 =0 )=\frac{\pi \sin \mu}{2 \mu (\pi - \mu )} ,
\ee
found in \cite{young}. Here, $\mu = \cos ^{-1} \Delta$.      

 Finally, since we are truncating the Hilbert space it is important to estimate how good this approximation is.  
A standard measure of that is the truncation error defined as the sum of the eigenvalues of the density matrix
that are discarded.       
Keeping up to $32$ eigenstates of the density matrix we obtained a truncation 
error less than $10^{-7}$ for open boundary conditions (spin correlations) and less than $10^{-5}$ 
for twisted boundary conditions (superfluid density).
       
\section{Results} \label{ressec}

\subsection{$\langle S_i^x S_j^x + S_i^y S_j^y \rangle$ correlation}

Before summarizing the correlation functions results, let us discuss what should be the 
behavior of the correlation length in a phase diagram like the one proposed in 
Fig. \ref{phasediag1}. 
As we said before, in the localized phase it is possible to define a correlation length, 
i.e. the correlations functions display an exponential behavior at large distances. Instead, 
in the phase with quasi-LRO this is not possible 
because, since the region is critical, the correlation length is infinite. This means that, if we were able to 
solve the problem exactly, we should observe in a 3-dimensional plot of correlation length vs 
($\Delta$,$h_0$) that this quantity diverges when we approach the line that divides the two regimes. 
This is not the case when we consider a finite chain and moreover when we truncate the 
Hilbert space. What we should obtain instead is that the correlation length has a peak located in a region that 
coincides approximately with the critical region. This peak should become sharper, 
higher and closer to the transition as we consider larger chains and keep more DMRG states.           
 
Having this in mind, we computed the spin-spin correlations 
\be
\overline{C_t(L)}=\overline{\ln |\langle S_0^x S_L^x + S_0^y S_L^y\rangle|},
\ee
transverse to the random field for several points of the phase diagram $\Delta -h_0$. Here, the bar indicates 
average over disorder.

\begin{figure}
\centering
\epsfysize=8.0truecm
\epsffile{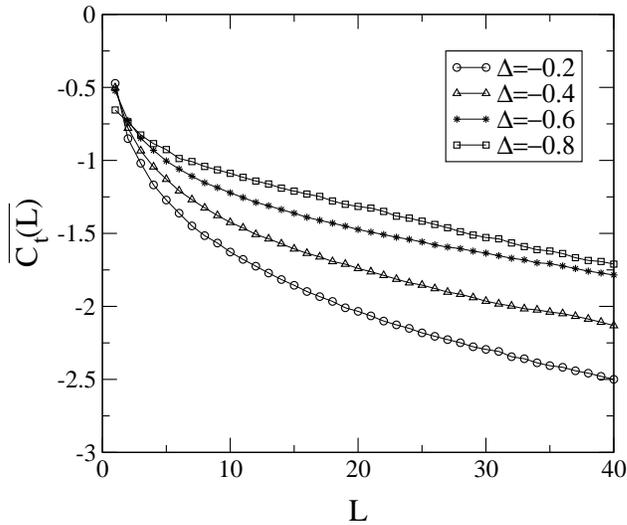}
\caption{Average of the logarithm of the correlations (typical correlations) vs. the distance for $h_0=0.3$ and different values of the anisotropy. The statistical error originated in the random average is, in all the cases, smaller 
than the size of the symbols.}
\label{logcor}
\end{figure}

In Fig. \ref{logcor}, $\overline{C_t(L)}$ is plotted for $h_0=0.3$ and different values of $\Delta$. 
The linear behavior at large distances means the correlations decay exponentially. 

\begin{figure}
\centering
\epsfysize=8.0truecm
\epsfbox{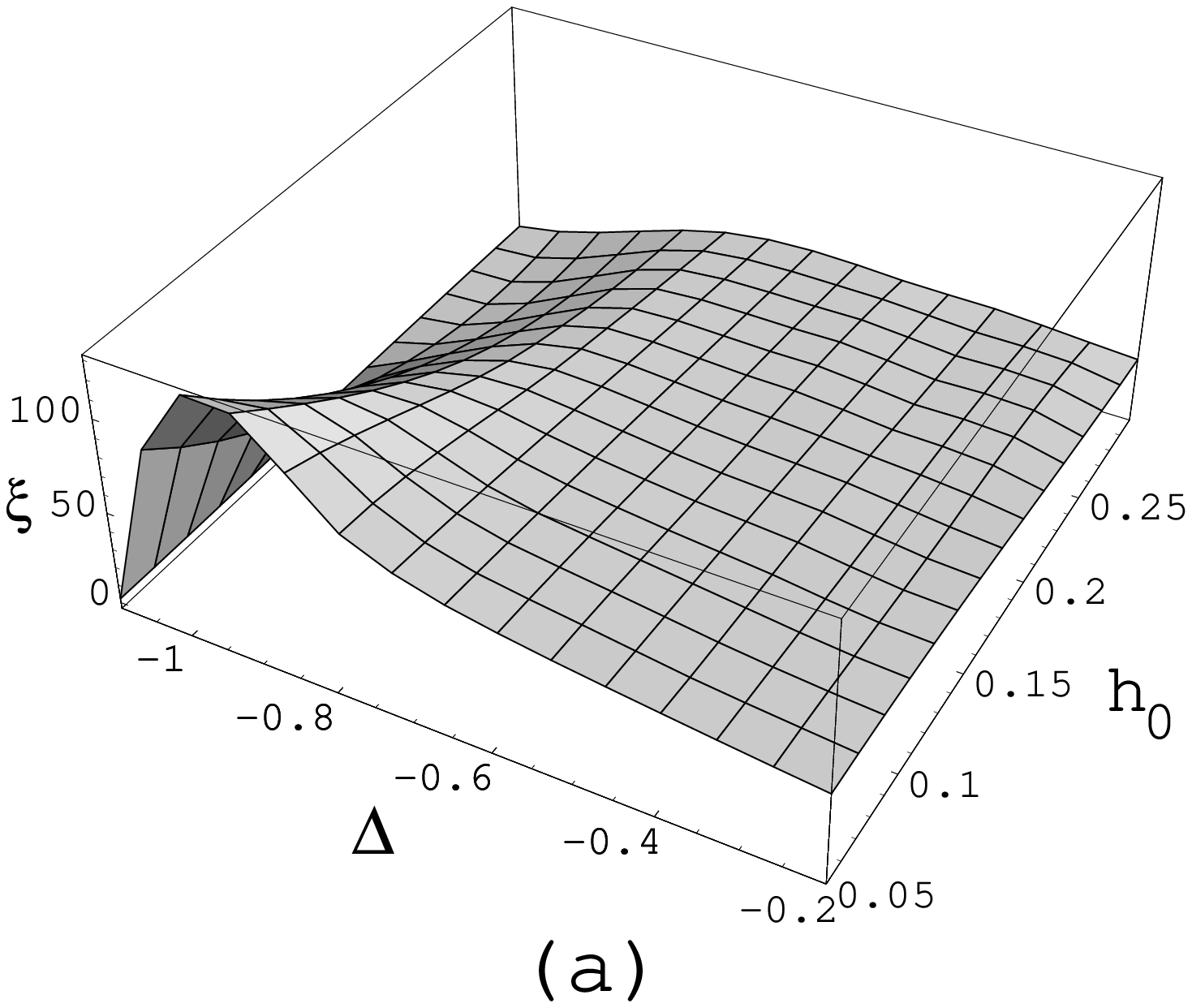}

\epsfysize=8.0truecm
\epsfbox{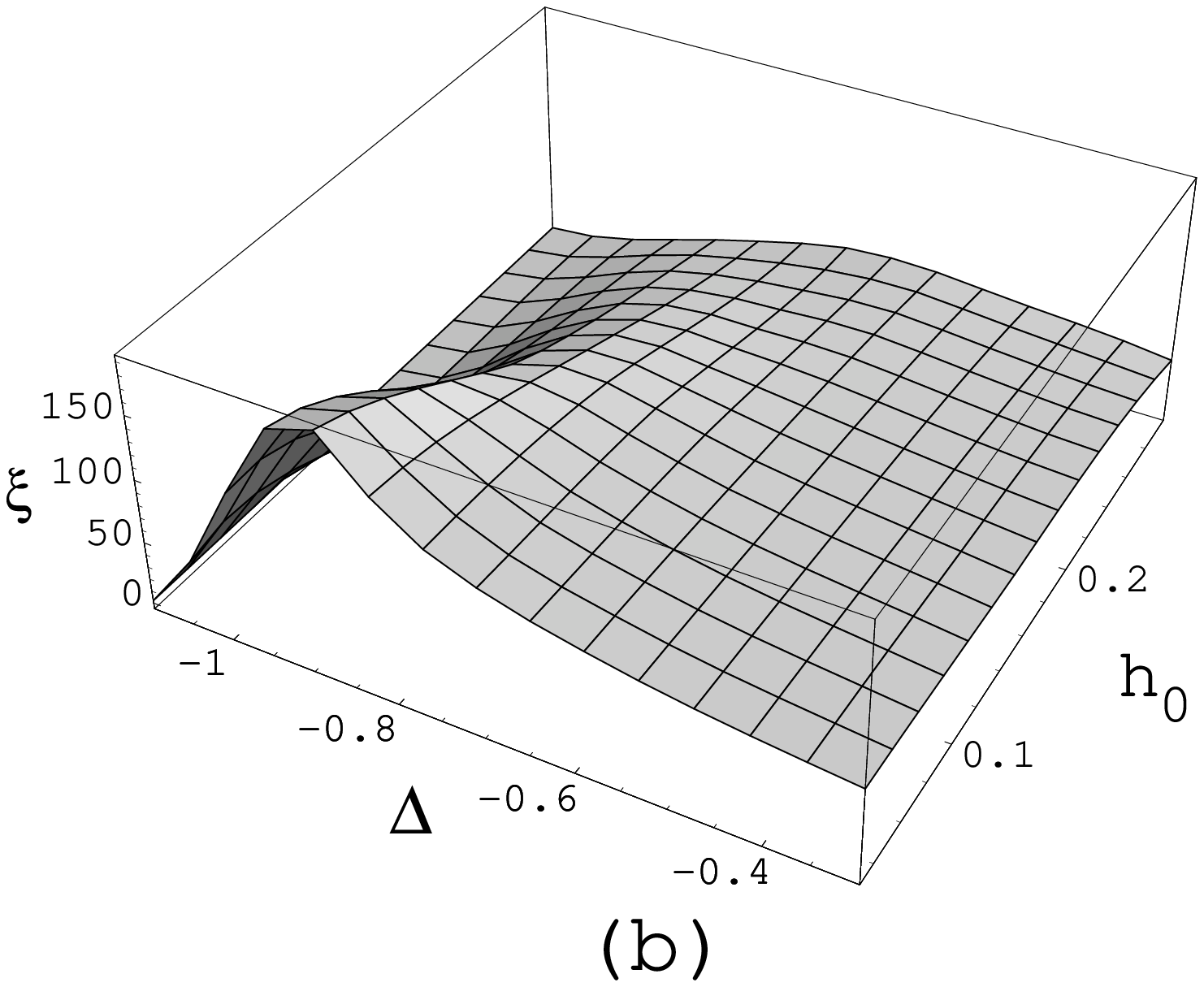}
\caption{The dependence of the typical correlation length in the plane $\Delta-h_0$. a) Keeping 16 eigenstates of the density matrix. 
b) Keeping 32 eigenstates of the density matrix. }
\label{cor3d}
\end{figure}

From the slope of these lines we estimate the typical correlation length $\xi$ and analyzing its behavior as a 
function of $\Delta$ and $h_0$ we obtain the 3-dimensional plot shown 
in Fig. \ref{cor3d} for $16$ and $32$ DMRG states. There, we can see that the correlation length not 
only increases when we approach the axis $h_0=0$ (pure system) keeping $\Delta$ constant but also in a region with finite 
disorder centralized in a negative value of $\Delta$ between $-1$ and $-0.5$. This 
is an indication of the presence of a critical region for finite disorder where the quasi-LRO 
found for $h_0=0$ persists. 

\begin{figure}
\centering
\epsfxsize=8.0truecm
\epsfbox{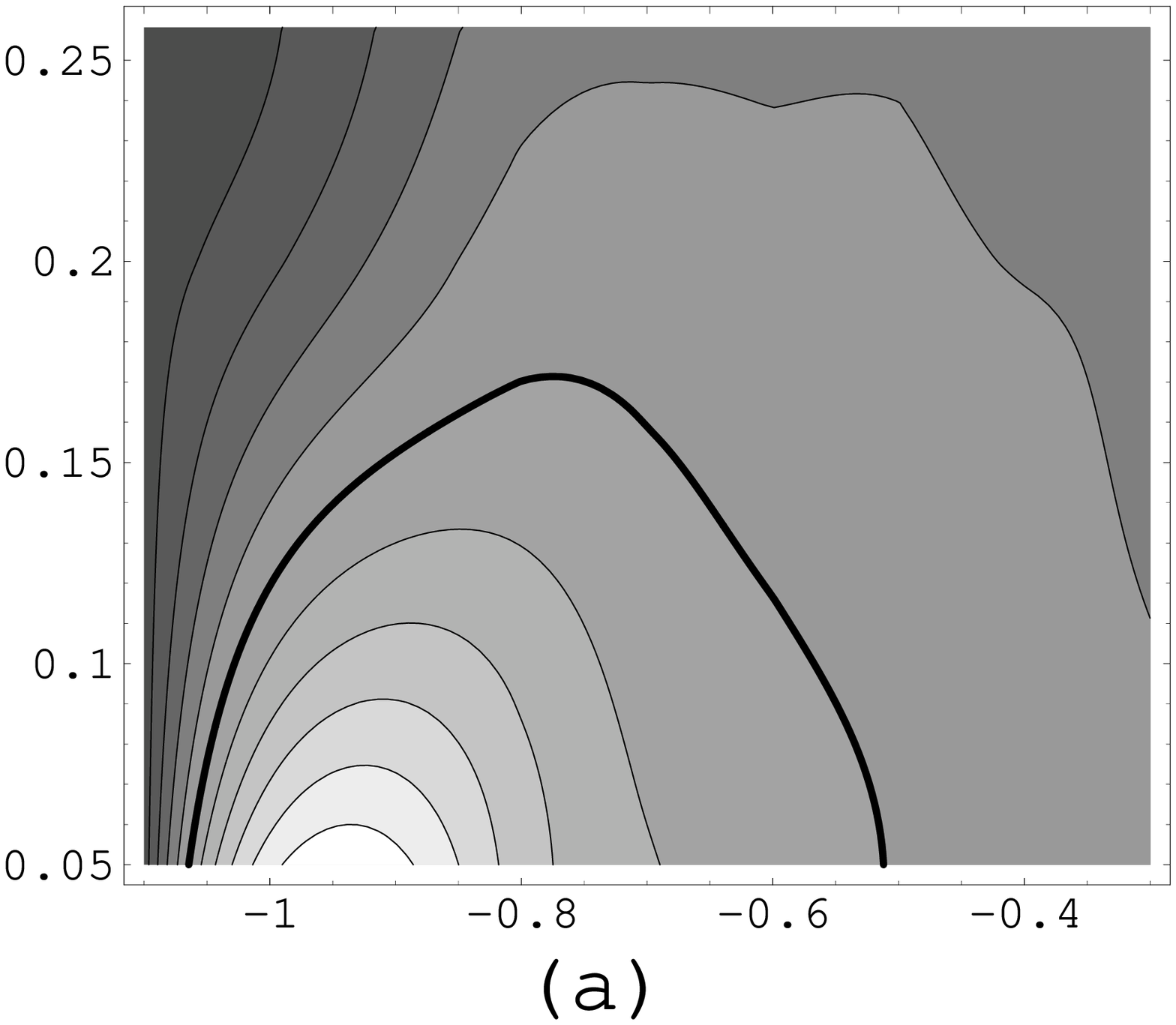}

\epsfxsize=8.0truecm
\epsfbox{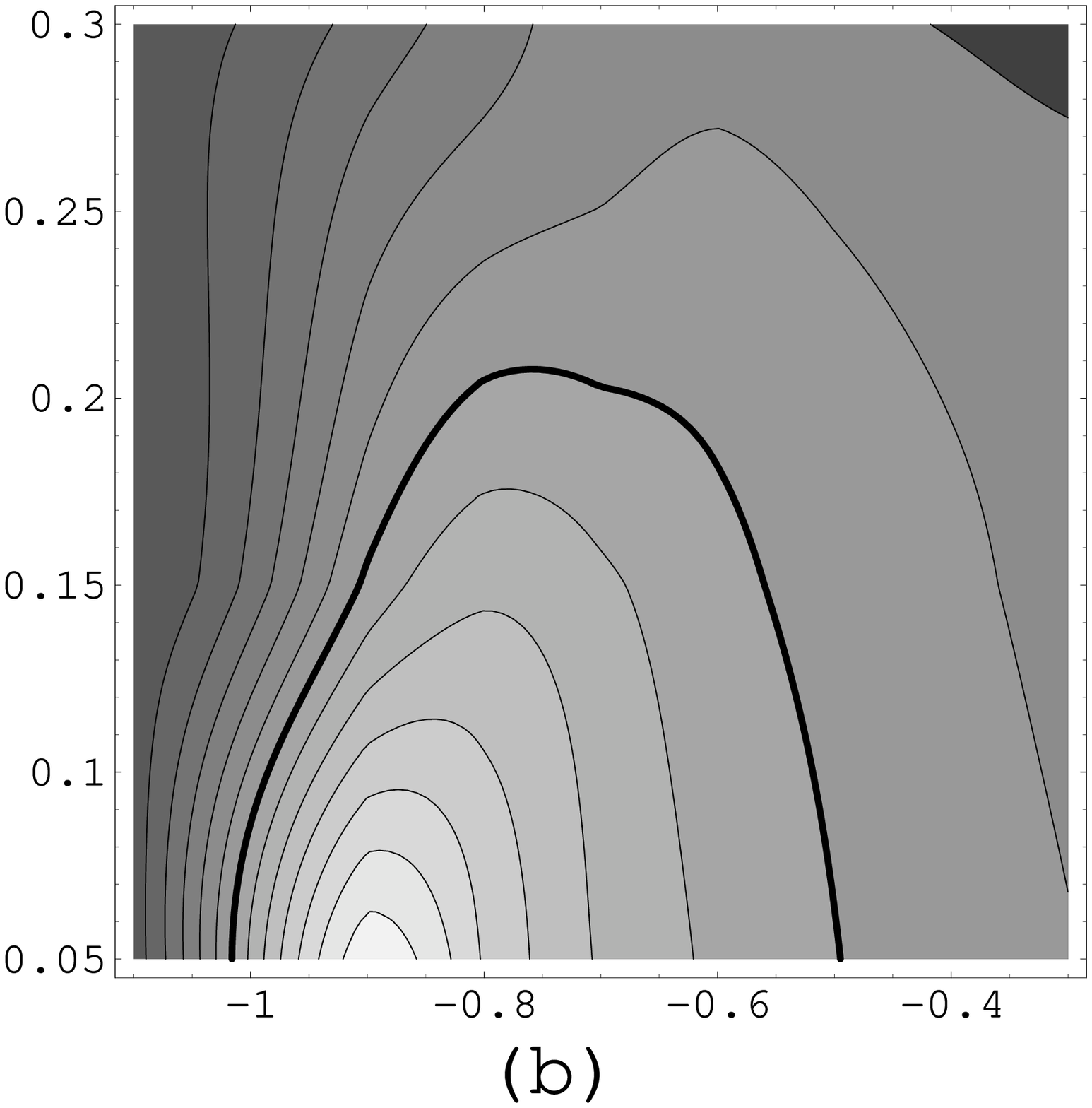}
\caption{The contour plot of the typical correlation length in the plane $\Delta-h_0$. a) Keeping 16 eigenstates of the density matrix. 
b) Keeping 32 eigenstates of the density matrix.}
\label{contour}
\end{figure}

To obtain an approximation to the shape of the critical region, we propose to analyze the contour plot of the typical correlation length.   
Following the previous discussion, in the limit of infinite length and infinite number of DMRG states, the contour lines will concentrate on the border of the critical region. If, for a given truncation, we use the input that the critical region starts at $\Delta=-0.5$, the choice of a contour line such that 
$h_0 \approx 0$ at $\Delta=-0.5$ will give us qualitative information about the shape of the critical region. This choice will be justified in the next section when we discuss the superfluid density results.        
Fig. \ref{contour} shows the contour plot of the typical correlation length for $16$ and $32$ DMRG states. From the chosen contour lines (bold lines in the figure) we estimate $ \hc (\Delta=-0.75) \simeq 0.17$ 
for $16$ states kept and $ \hc (\Delta=-0.75) \simeq 0.20$ keeping $32$ states.  
We want to remark here that in Fig. \ref{contour}, the contour lines used to estimate the boundary 
of the critical region correspond to values of $\xi$ smaller than 100 (see Fig \ref{cor3d}) while 
the chains considered were larger than 200 sites. Thus, the finite size effects are small and the main 
approximation is in the truncation of the Hilbert space.

\subsection{Superfluid density}

\begin{figure}
\centering
\epsfysize=8.0truecm
\epsffile{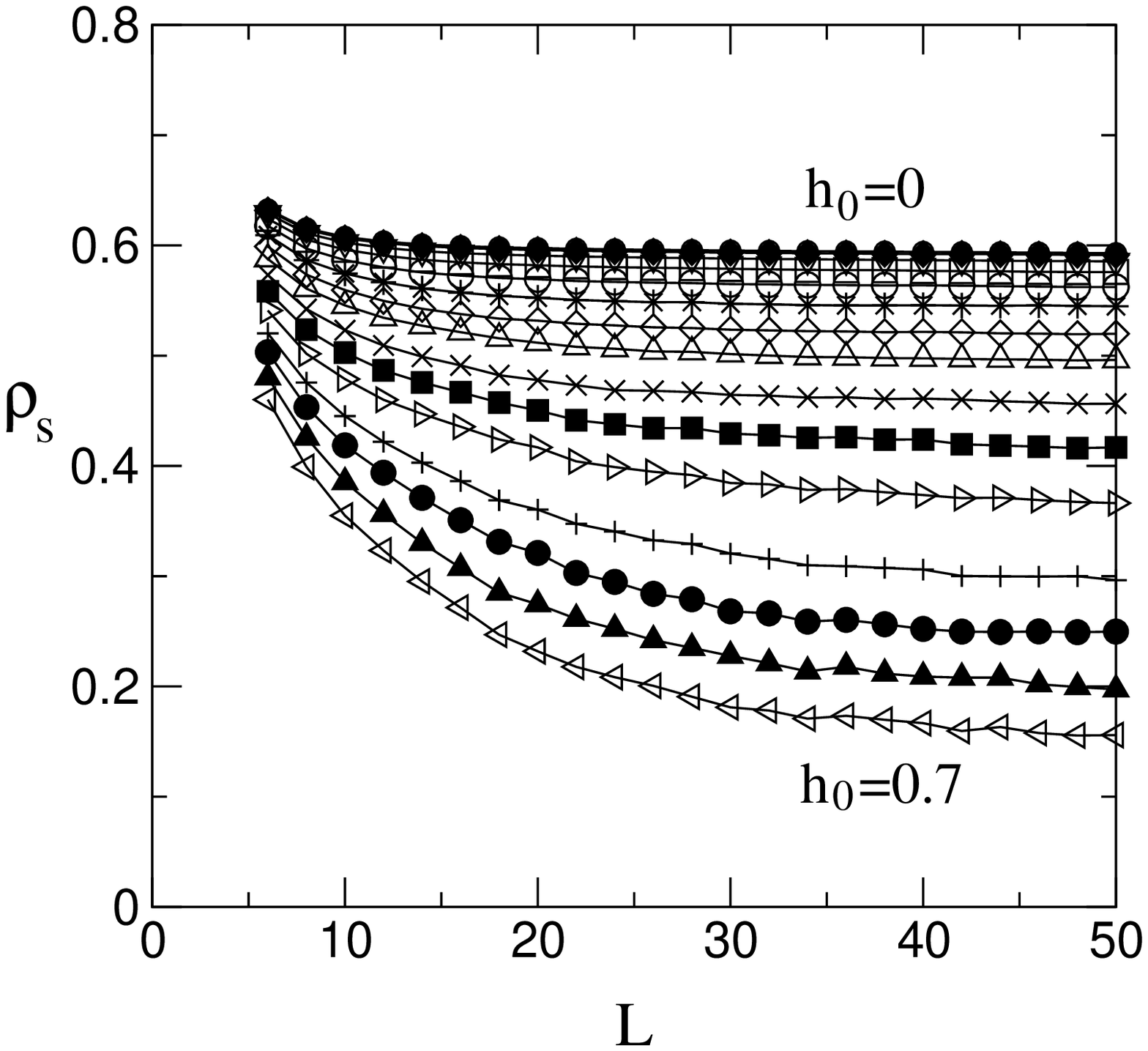}
\caption{The dependence of the superfluid density with the number of sites for different values of the random field and $\Delta=-0.75$. The statistical error originated in the random average is, in all the cases, smaller than the size of the symbols.}
\label{rhovsNdifh}
\end{figure}

We computed the superfluid density using (\ref{rhocal}) and performed afterwards the average over disorder. 
In Fig. \ref{rhovsNdifh}, we show $\rho_s$ vs $L$ for different values of the strength of the random field $h_0$ 
and $\Delta=-0.75$. As expected, for a fixed length of the chain, $\rho_s$ decreases with $h_0$. Note that we obtain 
non zero values of $\rho_s$ because we study finite systems. 
For the infinite system the superfluid density is different from zero only for ($h_0=0$,$|\Delta|<1$) and inside the 
critical region, that is, where the correlation length diverges. For that reason $\rho_s$ can be used to study the phase 
diagram of Fig. \ref{phasediag1}.

\begin{figure}
\centering
\epsfysize=8.0truecm
\epsffile{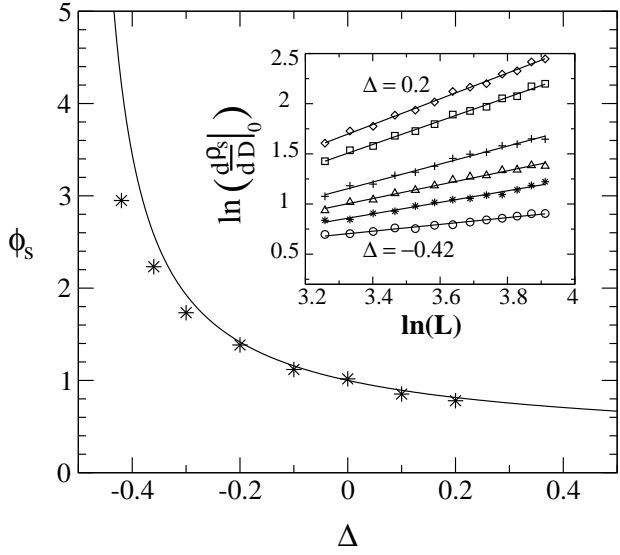}
\caption{The dependence of the correlation length critical exponent with the anisotropy. The stars correspond to 
our numerical data and the solid line is the perturbative renormalization group prediction \cite{fisher}. The inset shows the 
dependence of $\ln (\partial \rho _s / \partial D |_{D=0})$ with $\ln L$ for different anisotropy values; the slopes correspond 
to the inverse of the correlation length critical exponent.}
\label{fis}
\end{figure}

First we are going to check that the quasi-LRO region extends up to $\Delta=-0.5$. This point is characterized by the fact 
that there, the correlation length critical exponent diverges. In the region $-1/2< \Delta <1$ where 
the disorder is relevant, the dependence of the correlation length with the disorder $D = h_0^2/3$ near the critical 
line corresponding to $h_0=0$, should be $\xi \sim D^{- \phi _s}$ and according to the perturbative renormalization group 
results \cite{fisher}, the critical exponent $\phi_s$ diverges at $\Delta=-0.5$ like  
\be
\phi _s = \frac{1}{3-(1-\frac{1}{\pi} \arccos \Delta)^{-1}}.
\label{eq:fis}
\ee 
To extract this exponent from our numerical data, we analyzed the derivative $\partial \rho _s / \partial D$ at 
$D=0$. Assuming that $\rho_s = \rho_s (L/\xi)$ for $L \gg 1$ \cite{runge}, the behavior of 
$\ln (\partial \rho _s / \partial D |_{D=0})$ with $\ln L$ should be linear with the slope corresponding to $\phi _s ^{-1}$. In the inset of Fig. \ref{fis} we show this linear behavior for different values of 
$\Delta$.  
The critical exponent extracted from our data and the analytical prediction (eq. \ref{eq:fis}) are shown 
in Fig. \ref{fis} where a good agreement can be observed. This result is in line with the existence of a 
critical region for $\Delta < -0.5$ where the disorder is irrelevant and justify our choice made in the 
$\xi$ contour plots.      

\begin{figure}
\centering
\epsfysize=8.0truecm
\epsffile{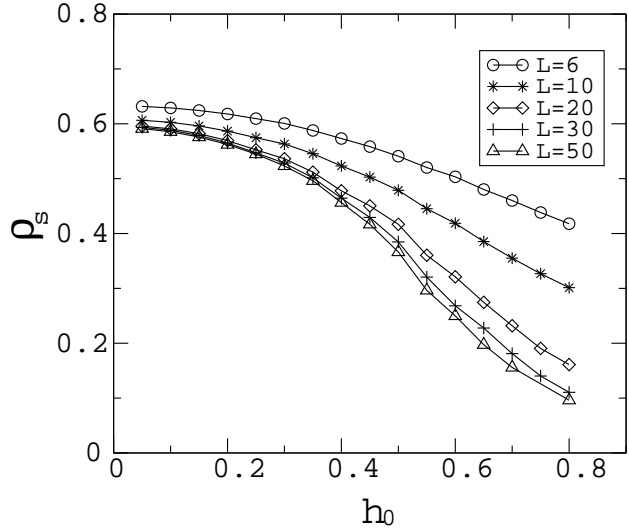}
\caption{The dependence of the superfluid density with the strength of the random field for systems with different number of sites. The the anisotropy value is $\Delta = -0.75$.}
\label{rhovshdifN}
\end{figure}

For a given truncation, the border of this region can be estimated studying the dependence of $\rho_s$ 
with the strength of the 
random field for different lengths of the chain, $L$. 
Figure \ref{rhovshdifN} is a plot of $\rho_s$ vs $h_0$ for $\Delta=-0.75$ and different lengths. 
The exact solution (infinite length and infinite DMRG states) should show a 
discontinuity in a transition of the Kosterlitz-Thouless 
type ($\rho_s(h_0 < \hc) > 0$ 
 and $\rho_s(h_0 > \hc)= 0$).

\begin{figure}
\centering
\epsfysize=8.0truecm
\epsffile{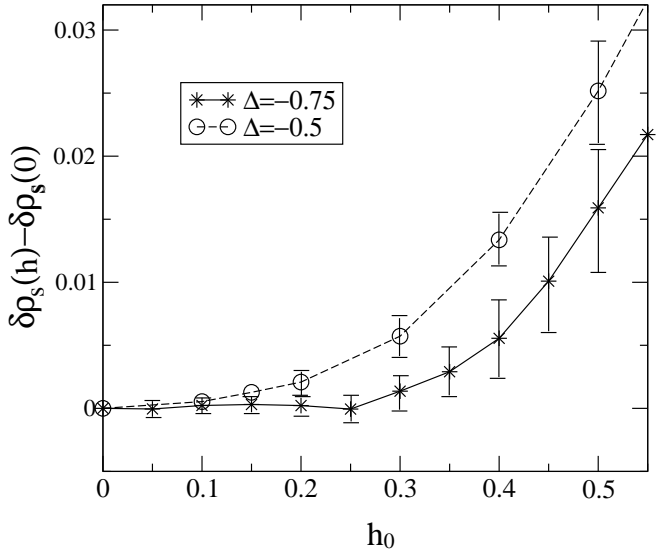}
\caption{The dependence of $\delta \rho_s$ with $h_0$ for $\Delta=-0.5$ and $\Delta=-0.75$. The error bars indicate 
the statistical error originated in the random average.}
\label{deltarho1}
\end{figure}

Using this fact, an estimation of the critical field can be obtained analyzing the behavior of the 
function $\delta \rho_s (\Delta ,h_0)= \rho_s(\Delta ,h_0,L)- \rho_s(\Delta ,h_0,L')$ where $L$ 
and $L'$ are fixed and $L,L'\gg 1$. 
This corresponds, in Fig. \ref{rhovshdifN}, to taking two of the curves and 
computing the difference. If the two curves correspond to lengths that are $L,L' \gg 1$, the difference 
between them will start to increase considerably only when $h_0 > \hc$. This criterion assumes 
that for $L \gg 1$ $\rho_s(L,\xi)=\rho_s(L/\xi)$ and then, if the truncated system correlation length is 
much larger than $L$, $\rho_s$ becomes almost independent on the length $L$. In this way,  
this estimation is affected mainly by the truncation of the DMRG method and then, we expect to have 
$\delta \rho_s$ almost constant in the region where the truncated system correlation length is much 
larger than $L=50$ (which is the maximum length considered in the superfluid density calculations).   
In Fig. \ref{deltarho1} we show $\delta \rho_s(\Delta ,h_0)$ vs. $h_0$ (taking 
$L=30$ and $L'=50$) for $\Delta=-0.5$ and $\Delta=-0.75$. The points in these curves are obtained by 
averaging over disorder and have a statistical error which is shown in the figure.  
As it is possible to observe, for $\Delta=-0.75$ the function stays almost constant until a value of 
$h_0 \sim 0.25$. We have checked that this behavior is the same taking different values of 
$L$ and $L'$.

\begin{figure}
\centering
\epsfysize=8.0truecm
\epsffile{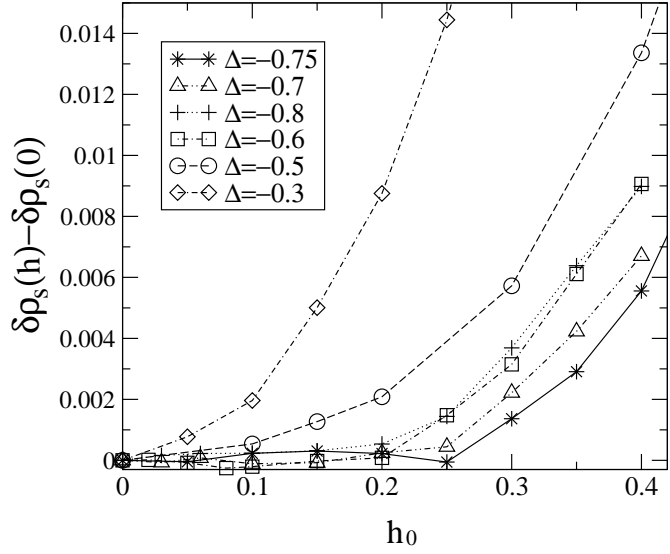}
\caption{$\delta \rho_s$ vs $h_0$ for different values of $\Delta$ keeping 32 eigenstates of the density matrix.}
\label{deltarho2}
\end{figure}

We also analyzed $\delta \rho_s (\Delta ,h_0)$ for other anisotropy values. 
The results are shown in Fig. \ref{deltarho2}. Remarkably, all the curves  
corresponding to values of $\Delta$  between $-1$ and $-0.5$ fall between those 
corresponding to $\Delta=-0.5$ and $\Delta=-0.75$ supporting the statement that 
the maximum critical field occurs for $\Delta=-0.75$. An estimation of the critical field is then given by the intersection of the curve with the $h_0$-axis. To obtain that intersection we use the first two points which differ from zero by more than 
the statistical error. Also from the error in these points we estimate the error in the critical field so computed.

\begin{figure}
\centering
\epsfysize=8.0truecm
\epsffile{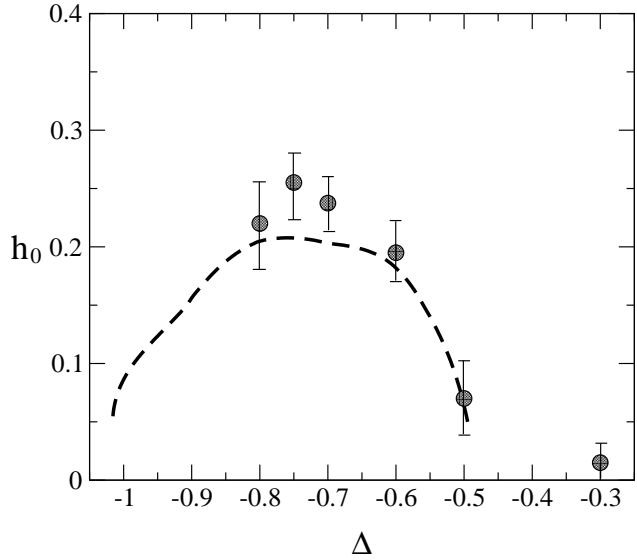}
\caption{Values of the critical field (keeping 32 states) obtained from Fig. \ref{deltarho2} 
as explained in the text. The dashed curve corresponds to the phase boundary estimated by the 
analysis of the correlations.}
\label{phasediag2}
\end{figure}

In Fig. \ref{phasediag2}, we plot these values of the critical field in the phase diagram giving 
an approximation to the boundary of the critical region. 

\section{Extrapolations for $\Delta=-0.75$} \label{extrapol}

In this section we study the special case of $\Delta=-0.75$ and 
perform the extrapolations to infinite length and infinite number of DMRG states of 
the computed properties for selected values of the random magnetic field. 

\begin{figure}
\centering
\epsfysize=8.0truecm
\epsfbox{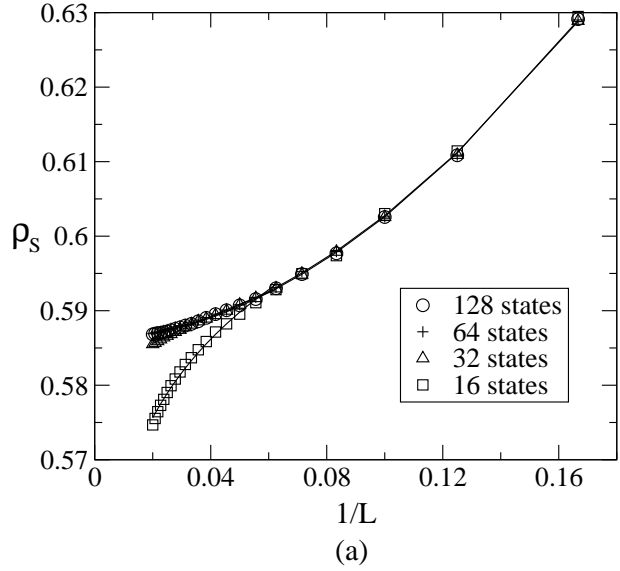}

\epsfysize=8.0truecm
\epsfbox{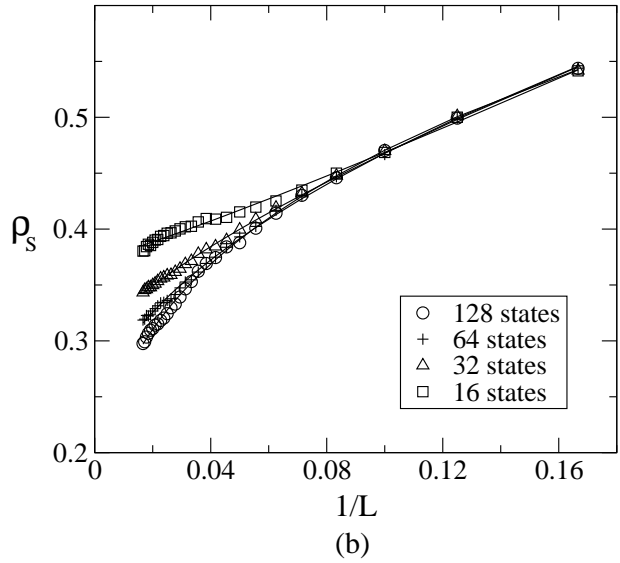}
\caption{The Superfluid Density for $\Delta=-0.75$ as a function of the inverse of the number of sites for 
different number of states kept in the DMRG. a) $h_0=0.1$. b) $h_0=0.5$. The solid lines are the fittings obtained 
using the expression $\rho_s (s,L) = \rho _0(s) + \alpha(s) /L^{\beta(s)} $. The statistical error originated 
in the random average is smaller than the size of the symbols.  At large $L$ the two cases display  
opposite behavior as we keep more DMRG states.}
\label{rhoh01h05}
\end{figure}

\begin{figure}
\centering
\epsfysize=8.0truecm
\epsffile{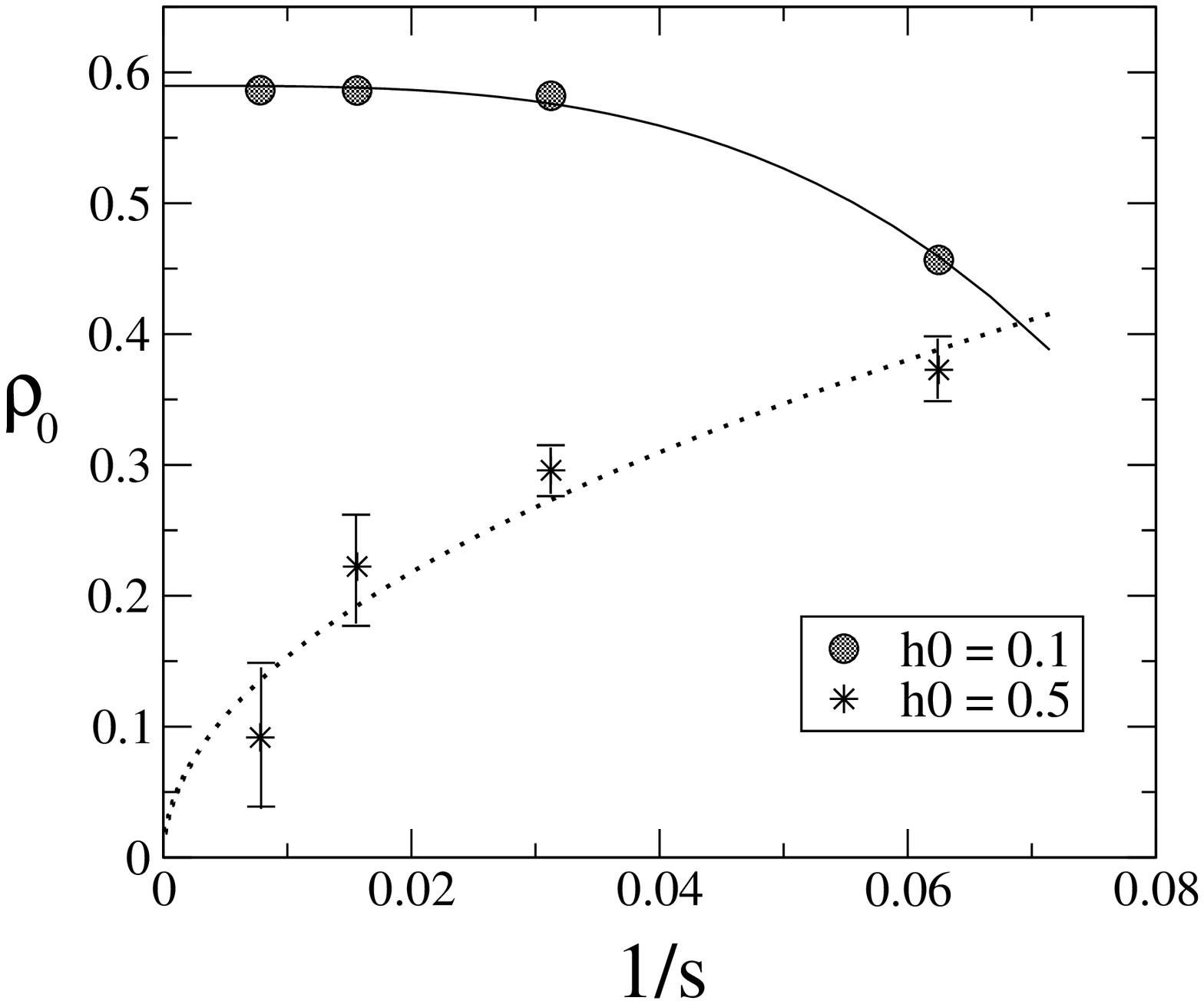}
\caption{Extrapolated values $\rho_0$ as a function of the inverse of the number of DMRG states. The lines are 
the result of the fittings using the expression $\rho_0 (s) = \rho_{\infty} + a /s^b $. For $h_0=0.1$ the 
extrapolated value is $\rho _{\infty} = 0.586 \pm 0.003$, whereas for $h_0=0.5$ the best fit 
was obtained for $\rho _{\infty} = 0$. The error bars were estimated from the statistical errors of 
the data. In the $h_0=0.1$ case they are smaller than the size of the symbols.}
\label{extrap}
\end{figure}

For the superfluid density $\rho_s$, we considered two points on the phase diagram: 
($\Delta = -0.75$, $h_0=0.1$) and ($\Delta = -0.75$, $h_0=0.5$). From the  
approximated phase diagram of Fig. \ref{phasediag2} we expect $h_0=0.1$ to be inside 
the critical region and $h_0=0.5$ to be outside.  
 In Fig. \ref{rhoh01h05} we plot $\rho_s$ as a function of $1/L$. At large $L$ the 
two cases display opposite behavior. For $h_0 =0.1$, $\rho _s$ increases as we keep 
more DMRG states whereas for $h_0=0.5$, it decreases. This is already an indication 
that the point ($\Delta = -0.75$, $h_0=0.1$) is inside the quasi-LRO region whereas 
the point ($\Delta = -0.75$, $h_0=0.5$) is outside. 
To find the asymptotic values  $\rho _0(s)=\rho _s (s,L\rightarrow\infty)$, we 
fitted each curve in Fig. \ref{rhoh01h05} using the expression 
$\rho_s (s,L) = \rho _0(s) + \alpha(s) /L^{\beta(s)} $. The solid lines in 
Fig. \ref{rhoh01h05} are the results of these fittings. Finally, the values of 
$\rho _0(s)$ vs. $1/s$ are plotted in Fig. \ref{extrap}. 
Again, we considered the expression $\rho_0 (s) = \rho_{\infty} + a /s^b $ to fit 
the curves. For $h_0=0.1$ the extrapolated value is $\rho _{\infty} = 0.586 \pm 0.003$,  
confirming that the point ($\Delta = -0.75$, $h_0=0.1$) is inside the 
critical region whereas for $h_0=0.5$ the best fit was obtained for $\rho _{\infty} = 0$, 
indicating the point ($\Delta = -0.75$, $h_0=0.5$) is in the localized phase as we 
expected from the previous analysis.

\begin{figure}
\centering
\epsfysize=6.6truecm
\epsfbox{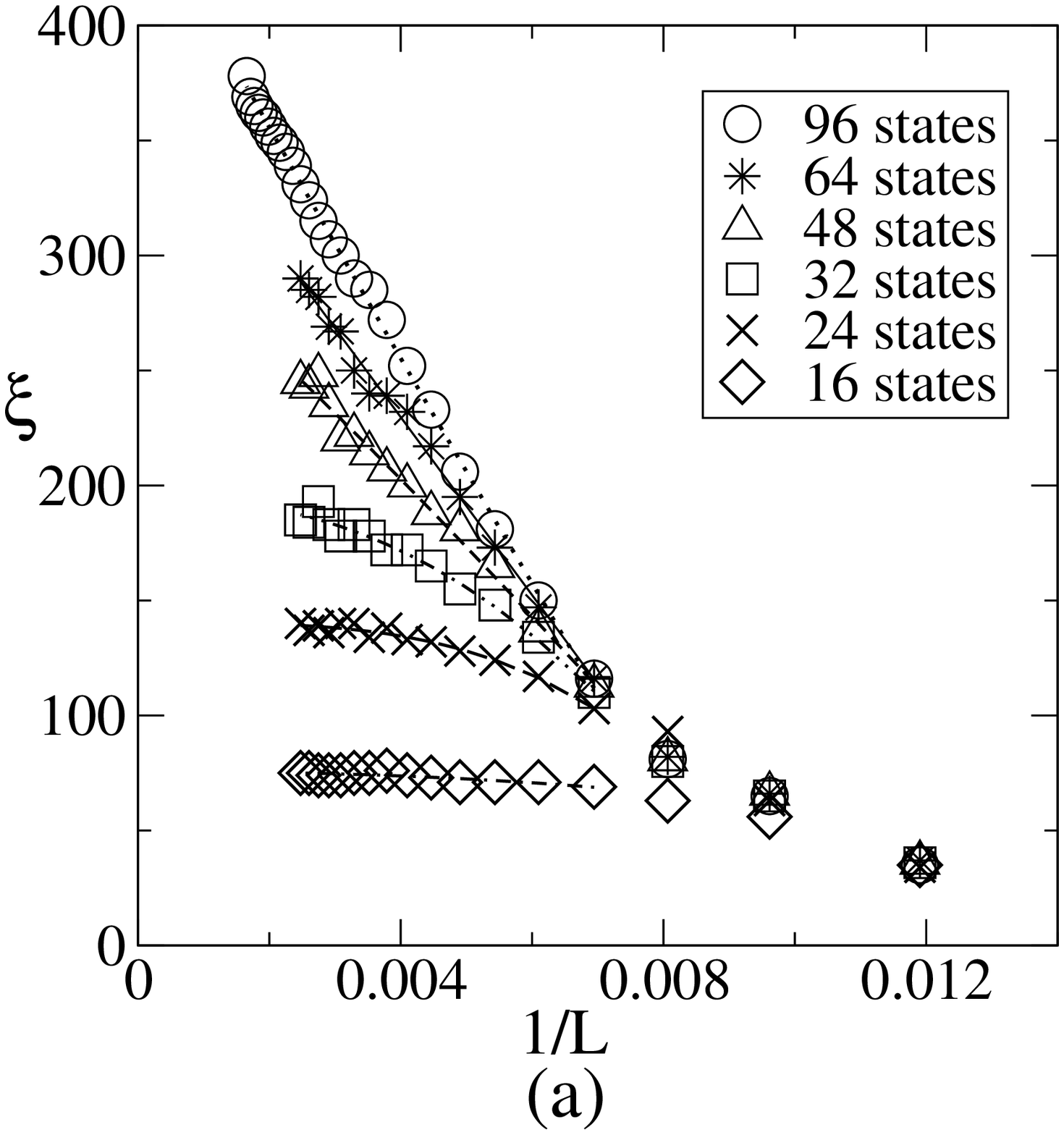}

\epsfysize=6.6truecm
\epsfbox{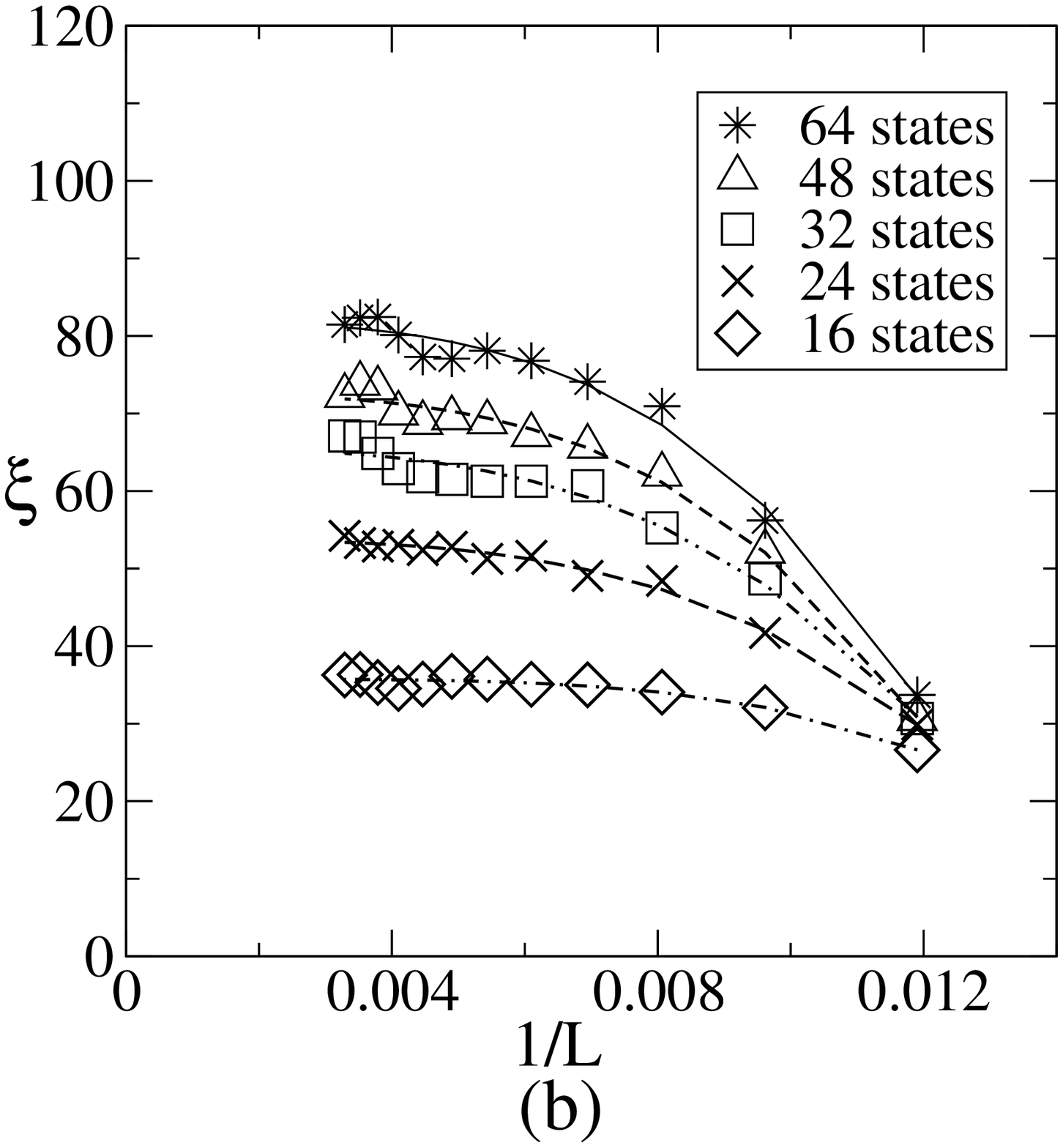}

\epsfysize=6.6truecm
\epsfbox{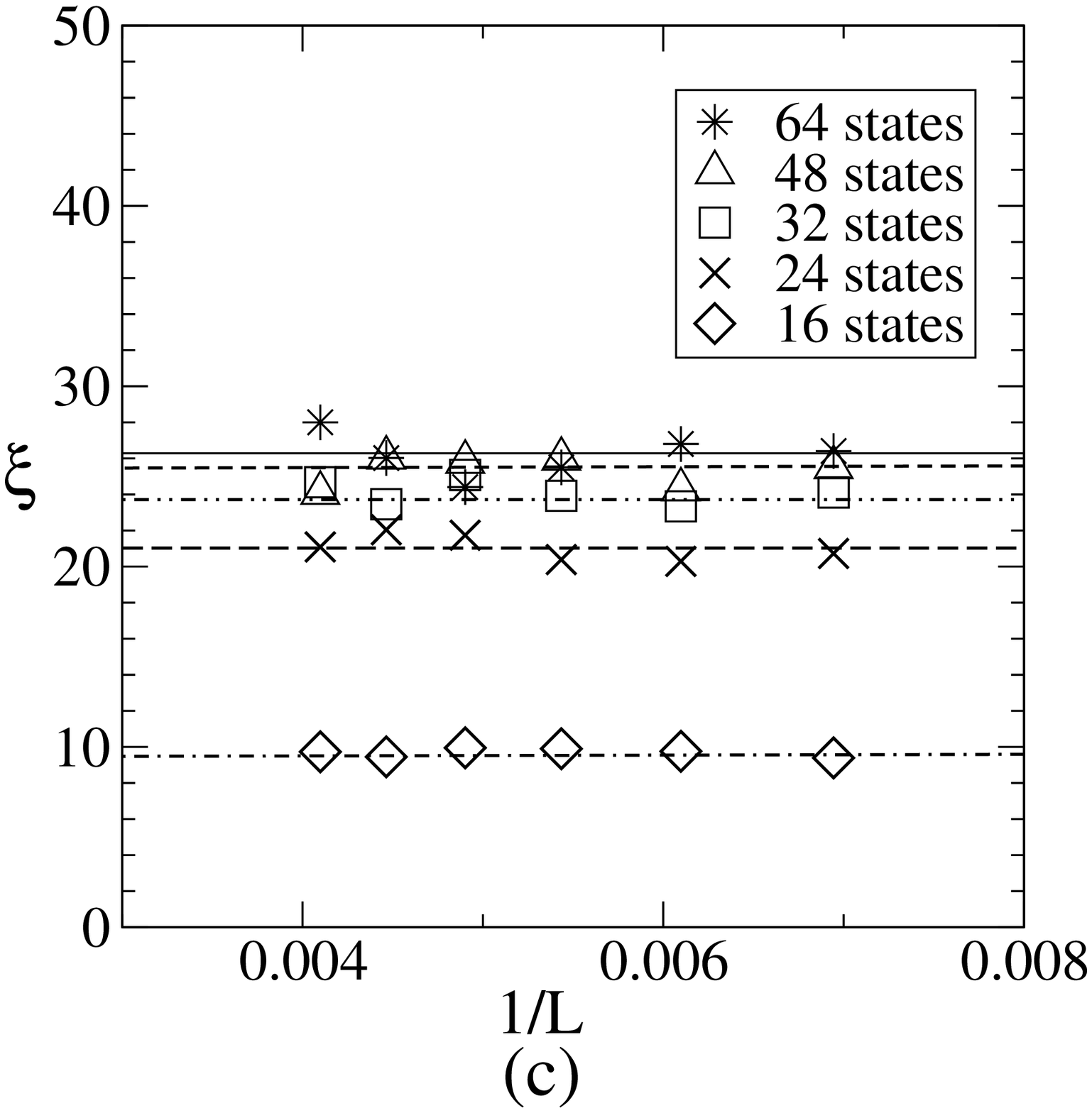}
\caption{The typical correlation length for $\Delta=-0.75$ as a function of the inverse of the number of sites for different number of states kept in the DMRG. a) $h_0=0.1$. b) $h_0=0.3$ c) $h_0=0.5$. The lines are the fittings obtained 
using the expression $\xi (s,L) = \xi _0(s) + \alpha(s) /L^{\beta(s)} $. The statistical error originated in the random average is smaller than the size of the symbols.}
\label{clevsL}
\end{figure}

\begin{figure}
\centering
\epsfysize=8.0truecm
\epsffile{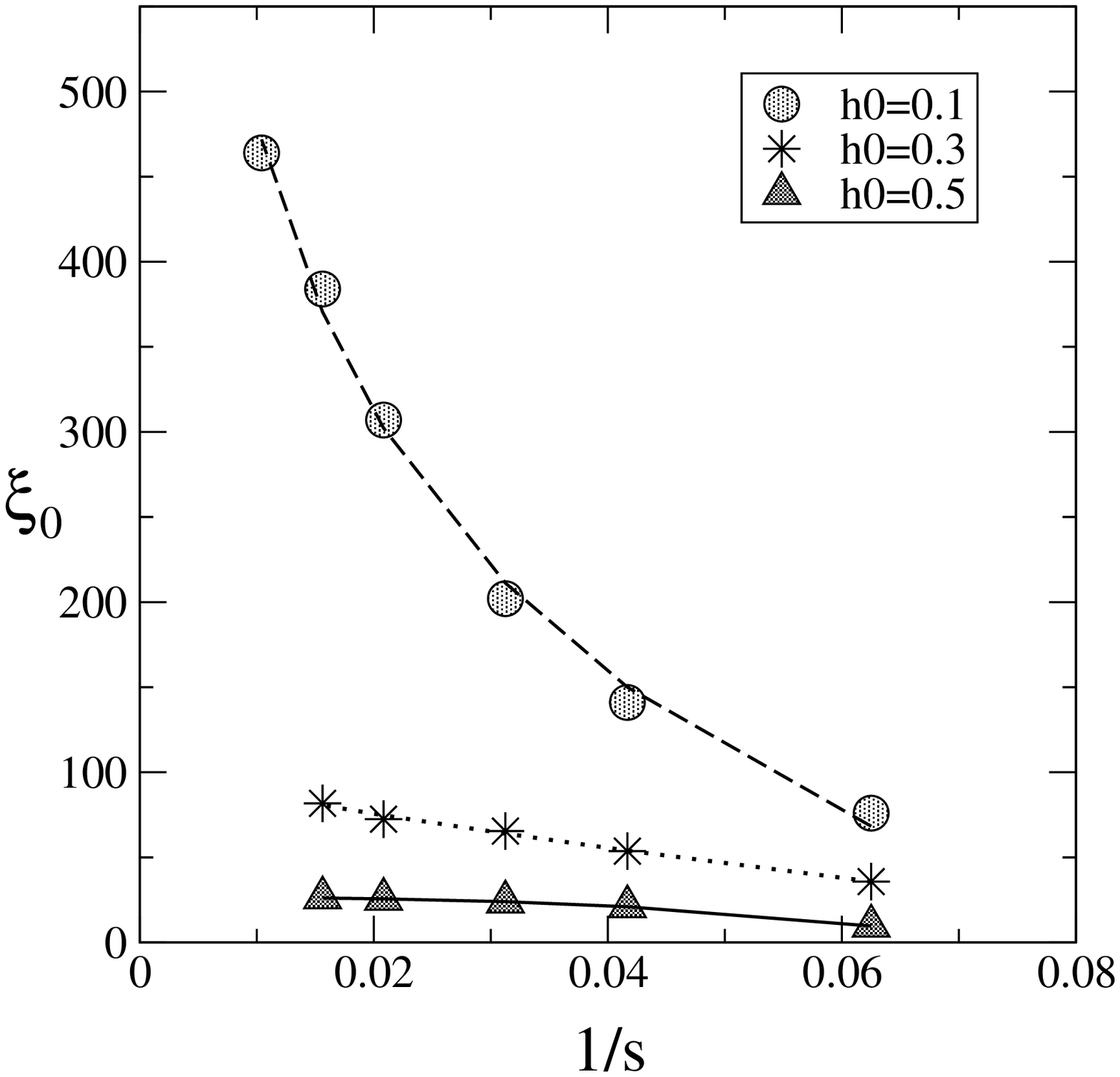}
\caption{Extrapolated values $\xi_0$ as a function of the inverse of the number of DMRG states. The lines are 
the result of the fittings using the expression $\xi_0 (s) = \xi_{\infty} + a /s^b $. 
The error bars are smaller than the size of the symbols.
For $h_0=0.1$ we find $b$ negative implying an infinite correlation length for infinite number 
of DMRG states. For $h_0=0.3$ and $h_0=0.5$ we find $b$ possitive and the extrapolated 
values for the correlation length are $\xi_{\infty} (h_0=0.3)= 108 \pm 18$ and 
$\xi_{\infty} (h_0=0.5)= 26.45 \pm 2.95$.}
\label{cleinfvsS}
\end{figure}

We proceeded in the same way with the correlation length. In Fig. \ref{clevsL} we 
plot the typical correlation length $\xi$ as a function of $1/L$ for 
$\Delta=-0.75$ and $h_0=0.1$, $h_0=0.3$ and $h_0=0.5$. In the $h_0=0.1$ case, the 
correlation length increases drastically as we keep more and more DMRG states. As 
more states are kept, the larger has to be the chain to reach the 
asymptotic value corresponding to the infinite system. Instead, for $h_0=0.3$ 
and $h_0=0.5$, the behavior is different. Not only the asymptotic values for the 
correlation length are one order of magnitude smaller but also 
the curves become closer as we increase the number of DMRG states 
indicating that the correlation length is finite. 
As we did for the superfluid density, we found the asymptotic values  
$\xi _0(s)=\xi (s,L\rightarrow\infty)$, fitting each curve in Fig. \ref{clevsL} 
with the expression $\xi (s,L) = \xi _0(s) + \alpha(s) /L^{\beta(s)} $. The lines in 
Fig. \ref{clevsL} are the results of these fittings. The asymptotic values 
$\xi _0(s)=\xi (s,L\rightarrow\infty)$ found in this way are plotted as a function 
of $1/s$ in Fig. \ref{cleinfvsS}. In all the cases, the errors in the data are 
smaller than $1\%$ and then, smaller than the size of the symbols. 

To fit the curves in Fig. \ref{cleinfvsS} we used again the expression 
$\xi_0 (s) = \xi_{\infty} + a /s^b $. 
For $h_0=0.1$ the best fit was obtained for a negative exponent ($b=-0.155$)  
implying an infinite correlation length for infinite number of DMRG states.   
On the other hand, for $h_0=0.3$ and $h_0=0.5$ we find $b$ possitive and 
the extrapolated values for the correlation length are 
$\xi_{\infty} (h_0=0.3)= 108 \pm 18$, and $\xi_{\infty} (h_0=0.5)= 26.45 \pm 2.95$.

In summary, from the extrapolations performed for $\Delta=-0.75$, we conclude the 
existence of a quasi-LRO and put bounds on the exact value of the critical field.  
This value should be larger than $h_0=0.1$ and smaller than $h_0=0.3$.

\section{Conclusions} \label{concsec}

In this work we analyzed numerically the quantum phase diagram of the spin $1/2$ XXZ chain in a random magnetic field pointing in the $z$ direction.

The two-point spin-spin correlation functions in the XY plane and the superfluid density were computed using an 
infinite-size density matrix renormalization group (DMRG) algorithm.

The asymptotic exponential behavior of the spin correlations allowed us to extract the typical correlation 
length and study its dependence with the anisotropy $\Delta$ and the strength of the random 
magnetic field $h_0$. The contour plots of the correlation length as a function of $\Delta$ and $h_0$  give numerical evidence of 
the existence of a critical region for finite disorder and negative values of the anisotropy. In this region the correlation length 
increases as we keep more states in the DMRG method. 
Using our data for $\rho_s$ we extracted the behavior of the correlation length critical exponent. A good agreement with the analytical 
prediction was found \cite{fisher} supporting the idea that the critical region starts in $\Delta=-0.5$. 
Then, an approximation to the boundary of the critical region can be obtained 
from the contour plots of the typical correlation length taking a contour line 
such that for $\Delta=-0.5$ the value of the field is 
$h_0 \simeq 0$. In this way, we estimated $\hc (\Delta=-0.75) \simeq 0.20$ for $32$ states 
kept.   
From the analysis of the function $\delta \rho_s (\Delta,h_0)= \rho_s(\Delta,h_0,L)- \rho_s(\Delta,h_0,L')$ for $L,L' \gg 1$, we 
obtained an independent estimation of the critical field $\hc(\Delta)$ for several 
values of $\Delta$ and drew the corresponding points of the line that separates the 
quasi-LRO regime from the localized one. 
We found that the largest value of the critical field  occurs at $\Delta=-0.75$ and 
is approximately $\hc(\Delta=-0.75)=0.25$ for $32$ states kept, in close agreement 
with the results obtained from the correlation length analysis. 
Since these results are approximated and are mainly affected by the truncation of 
the Hilbert space, to further support them, we extrapolated the typical correlation 
length and the superfluid density to infinite length and infinite number of DMRG 
states for the special case of anisotropy, $\Delta=-0.75$, and selected values of 
the random magnetic field. 
The extrapolations allow us to conclude the existence of a quasi-long-range 
order region and put bounds on the critical field. 
The range of possible values for the critical field ($0.1 < \hc < 0.3$) 
contains the approximated results before obtained.

\section*{Acknowledgments}
We are grateful to P. Henelius and M. Gulacsi for useful discussions. This work was supported by The Swedish Natural Science 
Research Council, and with computing resources by The Swedish Council for Planning and Coordination of Research (FRN) and 
Parallelldatorcentrum (PDC), Royal Institute of Technology, Sweden.

\end{document}